%% file: main.tex
\documentclass{article}
\input{packages.tex}
\input{macro.tex}

\title{Improving Confidence Estimation on Out-of-Domain Data\\for End-to-End Speech Recognition}
\name{Qiujia Li$^{1*}$, Yu Zhang$^2$, David Qiu$^2$, Yanzhang He$^2$, Liangliang Cao$^2$, Philip C. Woodland$^1$\thanks{$^*$Work was done while Li was a part-time student researcher at Google.}}
\address{$^1$ University of Cambridge, UK, $^2$ Google LLC, USA\\
\footnotesize{$^1$\texttt{\{ql264,pcw\}@eng.cam.ac.uk}, $^2$\texttt{\{ngyuzh,qdavid,yanzhanghe,llcao\}@google.com}}\vspace{-1em}}

\begin{document}
\ninept
\maketitle

\input{text/abs.tex}
\input{text/intro.tex}
\input{text/method.tex}
\input{text/setup.tex}
\input{text/exp.tex}
\input{text/conclusion.tex}


\clearpage
\section{References}
\vspace{-0.2em}
\begingroup
\renewcommand{\section}[2]{}
\bibliographystyle{IEEEbib}
\bibliography{refs}
\endgroup
\end{document}

%% file: packages.tex
\usepackage[preprint]{spconf}
\copyrightnotice{\copyright IEEE 2022}
\toappear{To appear in \emph{Proc. ICASSP 2022, May 22-27, 2022, Singapore}}

\usepackage{color,xcolor}
\usepackage{epsfig}
\usepackage{graphicx}

\usepackage{array}
\usepackage{booktabs}
\usepackage{colortbl}
\usepackage{multirow}
\usepackage{float}
\usepackage{caption}
\usepackage[labelformat=simple]{subcaption}

\usepackage{footnote}
\makesavenoteenv{tabular}
\makesavenoteenv{table}

\usepackage{amsmath,amsfonts,amssymb,bm}
\usepackage[super]{nth}

\usepackage{changepage}
\usepackage{extramarks}
\usepackage{fancyhdr}
\usepackage{lastpage}
\usepackage{setspace}
\usepackage{soul}
\usepackage{xspace}

\usepackage{url}
\usepackage{hyperref}
\hypersetup{colorlinks=True, urlcolor=black}
\usepackage[numbers,sort&compress]{natbib}
\usepackage{algorithm, algorithmic}
\usepackage{enumitem}
\usepackage{verbatim}
\usepackage{pifont}
\usepackage[acronyms]{glossaries}
\glsdisablehyper

%% file: macro.tex
\newcommand{\sect}[1]{Section~\ref{sec:#1}}
\newcommand{\sectdot}[1]{Sec.~\ref{sec:#1}}

\newcommand{\figdot}[1]{Fig.~\ref{fig:#1}}
\newcommand{\tbl}[1]{Table~\ref{tab:#1}}

\newcommand{\twotbl}[2]{Tables~\ref{tab:#1} and \ref{tab:#2}}

\newcommand{\ignore}[1]{}

\makeatletter
\DeclareRobustCommand\onedot{\futurelet\@let@token\@onedot}
\def\@onedot{\ifx\@let@token.\else.\null\fi\xspace}

\def\ie{\emph{i.e}\onedot}

\def\wrt{w.r.t\onedot}

\makeatother

\definecolor{MyDarkBlue}{rgb}{0,0.08,1}
\definecolor{MyDarkGreen}{rgb}{0.02,0.6,0.02}
\definecolor{MyDarkRed}{rgb}{0.8,0.02,0.02}
\definecolor{MyDarkOrange}{rgb}{0.40,0.2,0.02}
\definecolor{MyPurple}{RGB}{111,0,255}
\definecolor{MyRed}{rgb}{1.0,0.0,0.0}
\definecolor{MyGold}{rgb}{0.75,0.6,0.12}
\definecolor{MyDarkgray}{rgb}{0.66, 0.66, 0.66}

\definecolor{camblue}{HTML}{0072CF}
\definecolor{camred}{HTML}{D6083B}
\definecolor{camgreen}{HTML}{55A51C}

\def\presec{\vspace{-0.3em}}
\def\postsec{\vspace{-0.4em}}
\def\posttbl{\vspace{-0.5em}}

%% file: text/abs.tex
\begin{abstract}
As end-to-end automatic speech recognition (ASR) models reach promising performance, various downstream tasks rely on good confidence estimators for these systems. Recent research has shown that model-based confidence estimators have a significant advantage over using the output softmax probabilities. If the input data to the speech recogniser is from mismatched acoustic and linguistic conditions, the ASR performance and the corresponding confidence estimators may exhibit severe degradation. Since confidence models are often trained on the same in-domain data as the ASR, generalising to out-of-domain (OOD) scenarios is challenging. By keeping the ASR model untouched, this paper proposes two approaches to improve the model-based confidence estimators on OOD data: using pseudo transcriptions and an additional OOD language model. With an ASR model trained on LibriSpeech, experiments show that the proposed methods can greatly improve the confidence metrics on TED-LIUM and Switchboard datasets while preserving in-domain performance. Furthermore, the improved confidence estimators are better calibrated on OOD data and can provide a much more reliable criterion for data selection.
\end{abstract}

\begin{keywords}
confidence scores, end-to-end, automatic speech recognition, out-of-domain
\end{keywords}

%% file: text/intro.tex
\section{Introduction}
\postsec
Confidence scores are an important attribute associated with speech recognisers~\cite{Wessel2001ConfidenceMF,Jiang2005ConfidenceMF,Yu2011CalibrationOC}. Various downstream tasks rely on high-quality confidence scores, such as keyword spotting, dialogue systems and active / semi-supervised learning~\cite{Chan2004ImprovingBN,Tr2005CombiningAA,Riccardi2005ActiveLT}. For hidden Markov model (HMM)-based systems, word posterior probabilities from lattices or confusion networks can provide reasonably good estimates of confidence scores~\cite{Evermann2000PosteriorPD,Mangu2000FindingCI}. In order to have more reliable estimates, many model-based approaches have been proposed~\cite{Seigel2011CombiningIS,Kalgaonkar2015EstimatingCS,DelAgua2018SpeakerAdaptedCM,Ragni2018ConfidenceEA,Li2019BidirectionalLR}. More recently, as end-to-end speech recognition systems reach competitive performance with a simplified pipeline~\cite{Li2021ABA,Tuske2021OnTL,Li2020DevelopingRM}, confidence estimation for end-to-end models has become a much-needed component of an ASR system.

Because popular end-to-end systems such as recurrent neural network transducers~\cite{Graves2012SequenceTW} and attention-based encoder-decoder models~\cite{Chorowski2015AttentionBasedMF} have auto-regressive decoders that depend on the full history, it is not straightforward to generate lattice-like representations for a large number of hypotheses, from which word posteriors can be derived~\cite{Oneata2021AnEO}. Recent work has focused on data-driven approaches, where dedicated neural networks are used to predict confidence scores with various features extracted from end-to-end ASR models. For example, multi-layer perceptrons (MLPs), recurrent neural networks (RNNs) and self-attention networks have been used to learn token-level~\cite{Woodward2020ConfidenceMI,Li2021ConfidenceEF}, word-level~\cite{Qiu2021LearningWC}, and utterance-level confidence scores~\cite{Kumar2020UtteranceCM,Li2021ResidualEM,Liu2021UtterancelevelNC}. \cite{Qiu2021MultiTaskLF} proposes to jointly learn word and utterance level confidence scores together with deletion prediction via multi-task learning. With effective confidence estimation, simply rescoring $n$-best hypotheses with confidence scores can directly improve the ASR performance~\cite{Li2021ResidualEM,Qiu2021MultiTaskLF}.

Although model-based confidence estimators can yield good performance on both HMM-based systems and end-to-end systems, they are normally trained on the same data as the ASR system. Therefore, it is questionable whether the confidence estimation on out-of-domain (OOD) data will be reliable~\cite{Ovadia2019CanYT}. Under mismatched acoustic and / or linguistic conditions, it may be hard for the ASR model to generalise well on unseen data. However, a reliable confidence estimator should ideally provide a good indication of the quality of the automatic transcription, even for OOD data. In this paper, assuming that the accessible OOD data is not transcribed, \ie only acoustic data and text data but they are not paired, and the ASR model is fixed, two approaches are proposed to improve OOD confidence estimation. By using automatically generated ``pseudo'' transcriptions on OOD acoustic data and features from an additional language model (LM) trained from OOD text, experiments show that the quality of confidence scores can be improved greatly on two OOD datasets while maintaining in-domain performance.

In the rest of the paper, \sectdot{cem} describes the model-based confidence estimator used for end-to-end ASR model and \sectdot{ood} presents the two approaches for improving OOD confidence scores. \sect{setup} details the data, model and metrics used for producing the experimental results in \sectdot{exp}. Conclusions are given in \sectdot{conclusion}.

%% file: text/method.tex
\section{Confidence Estimation for End-to-End ASR}
\postsec
\label{sec:cem}
In the confidence estimation module (CEM) approach~\cite{Li2021ConfidenceEF}, a small additional network that processes relevant features from attention-based encoder-decoder models is used to estimate the confidence scores for each output token. The CEM can improve the confidence estimation performance regardless of the regularisation techniques used during ASR training.

The residual energy-based model (R-EBM)~\cite{Li2021ResidualEM} was proposed to complement the locally normalised auto-regressive decoder to improve the recognition performance by $n$-best rescoring. Furthermore, R-EBMs can also be viewed as a discriminator between correct and erroneous hypotheses, which generates utterance-level confidence scores. Compared with simply aggregating all token-level confidence scores for an utterance-level score, directly optimising at the utterance level was shown to be better. Since the R-EBM is trained to automatically aggregate confidence information across the whole utterance, it can also take deletion errors into account.

\begin{figure}[t]
    \centering
    \includegraphics[width=\linewidth]{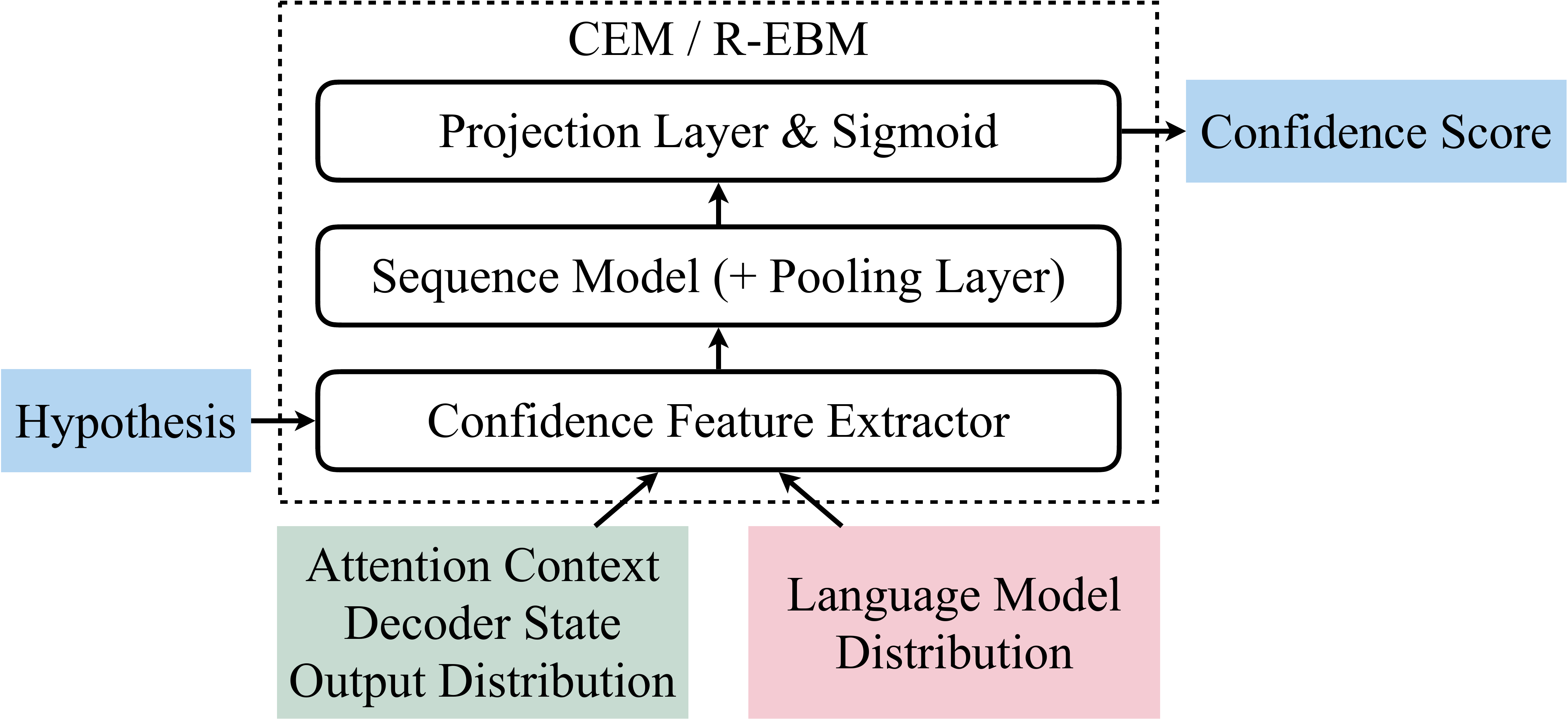}
    \caption{(Best viewed in color) The system schematic of CEM / R-EBM for confidence estimation. The pooling layer is only used in R-EBM. }
    \label{fig:system}
    \vspace{-1em}
\end{figure}

As shown in the dotted box in \figdot{system}, the CEM and R-EBM share the same skeleton that assigns the associated confidence scores for each hypothesis (\textcolor{camblue}{blue} blocks). They first gather various useful features for the encoder and the decoder for each token in the hypothesis, including the attention context, the decoder state, and the output distribution (the \textcolor{camgreen}{green} block). Optionally, the corresponding output distribution from an LM can also be included as additional features (\textcolor{camred}{red} block). These features are then passed to a sequence model such as a recurrent neural network or a self-attention network. For the CEM, the hidden representation for each token is projected to a scalar and then mapped to the confidence score between 0 and 1 by the Sigmoid function. For the R-EBM, a pooling layer reduces the hidden representations of the entire sequence to a single representation. A projection layer with a Sigmoid activation produces utterance-level confidence scores. During training, $n$-best hypotheses are generated from the fixed ASR model and the training targets (1 for a correct hypothesis and 0 otherwise) are obtained by aligning hypotheses with the ground truth transcription using an edit distance.

\section{Out-of-Domain Confidence Estimation}
\postsec
\label{sec:ood}
Using the CEM and R-EBM can provide much more reliable confidence scores than the softmax scores obtained directly from the ASR model. Although training data is augmented more aggressively during CEM / R-EBM training than ASR training to avoid being over-confident~\cite{Li2021ConfidenceEF,Li2021ResidualEM}, training data used for confidence estimation is often the same as the ASR model. A useful confidence estimator should not only perform well on in-domain data, but also generalise well to OOD data without modifying the ASR model. Assuming that some unlabelled OOD data is available, the following two methods are proposed to improve the OOD confidence scores: pseudo transcriptions and an additional language model.

\presec
\subsection{Pseudo Transcriptions}
\postsec
One approach of exposing the confidence estimators to OOD acoustic data is to include it in the training process (\textcolor{camblue}{blue} blocks in \figdot{system}). To this end, the existing ASR model can be used to transcribe unlabelled OOD data to give pseudo transcriptions without any data augmentation. During CEM / R-EBM training, $n$-best hypotheses are generated on-the-fly with data augmentation, and are then aligned with the pseudo transcription to produce binary confidence targets. Note that $n$-best hypotheses are nearly always erroneous \wrt the pseudo transcription because beam search is run with augmented input acoustic data. The effectiveness of this approach depends on the similarity between the in-domain and OOD data. If the OOD data has, for example, very mismatched acoustic conditions and a different speaking style, the quality of pseudo transcriptions may be poor, which gives misleading training labels for confidence estimators.

\presec
\subsection{Additional Language Model}
\postsec
Linguistic information can be very useful for confidence estimation. In CEM / R-EBM, an in-domain language model (LM) is normally used. However, the difference in speaking style can lead to very different linguistic patterns. For example, comparing audiobooks with telephone conversations, the vocabularies used are distinct, as telephone conversations are generally much more casual and spontaneous. Therefore, it may be useful to leverage the OOD text data to train an additional LM, which can provide additional features for confidence estimators (\textcolor{camred}{red} block in \figdot{system}). If either an in-domain LM or an OOD LM has a high probability for a hypothesis token, then the token is more likely to be recognised correctly.

%% file: text/setup.tex
\section{Experimental Setup}
\label{sec:setup}
\presec
\subsection{Data}
\postsec
The in-domain speech data used in this paper is from audiobooks. For training, there are 57.7 thousand hours of unlabelled speech data from Libri-light~\cite{Kahn2020LibriLightAB} (``unlab-60k'' subset) for unsupervised pre-training and 100 hours of transcribed data from LibriSpeech~\cite{Panayotov2015LibrispeechAA} (``train-clean-100'' subset) for fine-tuning. The text data for language modelling has around 810 million words. The standard dev and test sets from LibriSpeech~\cite{Panayotov2015LibrispeechAA} (``dev-clean/-other'' and ``test-clean/-other'') are used for in-domain development and evaluation.

Two out-of-domain datasets are used in the experiments. The TED-LIUM release 3 corpus~\cite{hernandez2018tedlium3} contains 452 hours of talks, which is in a slightly different domain from audiobooks. Although the talks are mostly well prepared, the speaking style is more casual and the language is sometimes colloquial. The text data to train language models has around 255 million words~\cite{Rousseau2014EnhancingTT}. The standard dev and test sets from TED-LIUM are used.

Another out-of-domain dataset is Switchboard (SWBD). It has 319 hours of telephony conversations for training. The text data for language modelling is the combination of SWBD transcriptions and Fisher Transcriptions, which has approximately 25 million words. The Hub5’00 set is used as the development set, and RT03 is used as the evaluation set\footnote{The LDC catalogue numbers are LDC97S62 for SWBD training set, LDC2002S09 and LDC2002T43 for Hub5’00, LDC2007S10 for RT03, and LDC2004T19 and LDC2005T19 for Fisher transcription.}. Since all SWBD data was collected at 8~kHz, it is upsampled to 16~kHz before sending to the ASR model.

\presec
\subsection{Models}
\postsec
The ASR model is an attention-based encoder-decoder network. The encoder architecture follows the ``Conformer XL'' setup in~\cite{Zhang2020PushingTL} with 24 Conformer layers and around 600M parameters. Wav2vec 2.0~\cite{Baevski2020wav2vec2A}, an unsupervised pre-training method, was used to pre-train the encoder using the unlabelled in-domain Libri-light data. The decoder is a single-layer long short-term memory (LSTM) network with 640 units. The randomly initialised decoder was fine-tuned jointly with the encoder initialised from wav2vec 2.0 using LibriSpeech 100h labelled data. Regularisation techniques such as SpecAugment~\cite{Park2019SpecAugmentAS}, dropout, label smoothing, Gaussian weight noise and exponential moving average were used to improve performance. The modelling units are a set of 1024 word-pieces~\cite{Schuster2012JapaneseAK} derived from the LibriSpeech 100h training transcriptions.

The confidence estimation models, \ie, CEM and R-EBM, were trained on the LibriSpeech 100h dataset while freezing the ASR model. 8-best hypotheses were generated on-the-fly with more aggressive SpecAugment masks to simulate errors on unseen data~\cite{Li2021ResidualEM}. Both the CEM and R-EBM have two-layer bi-directional LSTMs with 512 units in each direction. Since the direct output of the CEM is the token-level confidence, the word-level confidence is obtained by taking the minimum score among all the tokens for each word.

\presec
\subsection{Metrics}
\postsec
For ASR performance, the word error rate (WER) and sentence error rate (SER) are used. For confidence estimation performance, the area under the precision-recall curve (AUC) and equal error rate (EER) are reported for both word-level and utterance-level confidence scores. For a given threshold, precision is the fraction of true positives over all samples that are deemed to be positives by the confidence estimator, and recall is the fraction of true positives over all samples that are actually positive. A higher threshold generally implies a higher precision but a lower recall. Pairs of precision and recall values can be plotted by varying the threshold. The area under such a curve is an indication of the estimator's performance. Using precision-recall curves is more informative than the receiver operating characteristics (ROC) curves under imbalanced classes~\cite{Saito2015ThePP}. In practice, a downstream application normally needs to make decisions based on confidence scores. EER is where the false negative rate equals the false positive rate, which is the optimal operating point if false acceptance and false rejection have equal costs.

%% file: text/exp.tex
\section{Experimental Results}
\postsec
\label{sec:exp}
\presec
\subsection{Baselines}
\postsec
The simplest baseline for confidence estimation is to directly use the softmax probability from the decoder output as the confidence scores. As discussed in \cite{Li2021ConfidenceEF}, the softmax-based confidence scores can be severely impacted by the regularisation techniques used during training. As shown in \twotbl{word_baseline}{utt_baseline}, by having a dedicated confidence estimation module at either the word-level~\cite{Li2021ConfidenceEF} or utterance-level~\cite{Li2021ResidualEM}, the in-domain performance on LibriSpeech test-clean/-other is greatly improved. Especially at the utterance level, the AUC increases from around 70\% to around 90\% while the EER is approximately halved.
\begin{table}[ht]
    \centering
    \begin{tabular}{lrccccc}
        \toprule
        \multirow{2}{*}{dataset} & \multirow{2}{*}{WER} & \multicolumn{2}{c}{softmax} && \multicolumn{2}{c}{CEM~\cite{Li2021ConfidenceEF}} \\
        \cmidrule{3-4}\cmidrule{6-7}
         & & AUC$\uparrow$ & EER$\downarrow$ && AUC$\uparrow$ & EER$\downarrow$ \\
        \midrule
        LS (test-clean)  &  2.7 & 99.29 & 21.59 & & 99.64 & 16.40\\
        LS (test-other)  &  4.9 & 98.79 & 19.93 & & 99.40 & 15.39\\
        \midrule
        TED (test)    &  9.4 & 96.49 & 24.07 & & 98.78 & 18.28\\
        SWBD (RT03) & 28.3 & 92.88 & 20.15 & & 96.45 & 18.59\\
        \bottomrule
    \end{tabular}
    \caption{Baseline WERs (\%) and word-level confidence estimation performance in AUC (\%) and EER (\%).}
    \label{tab:word_baseline}
    \posttbl
\end{table}

\begin{table}[ht]
    \centering
    \begin{tabular}{lrccccc}
        \toprule
        \multirow{2}{*}{dataset} & \multirow{2}{*}{SER} & \multicolumn{2}{c}{softmax} && \multicolumn{2}{c}{R-EBM~\cite{Li2021ResidualEM}} \\
        \cmidrule{3-4}\cmidrule{6-7}
         & & AUC$\uparrow$ & EER$\downarrow$ && AUC$\uparrow$ & EER$\downarrow$ \\
        \midrule
        LS (test-clean)  & 31.4 & 77.63 & 41.97 & & 91.68 & 21.53\\
        LS (test-other)  & 45.2 & 67.84 & 38.63 & & 88.05 & 18.43\\
        \midrule
        TED (test)    & 72.1 & 33.57 & 47.66 & & 71.24 & 23.17\\
        SWBD (RT03) & 82.7 & 34.91 & 36.59 & & 52.31 & 28.35\\
        \bottomrule
    \end{tabular}
    \caption{Baseline SERs (\%) and utterance-level confidence estimation performance in AUC (\%) and EER (\%).}
    \label{tab:utt_baseline}
    \posttbl
\end{table}

On the two OOD sets, the benefit of using CEM or R-EBM is also noticeable. However, the relative improvement of confidence estimation performance by using a confidence module diminishes as the data becomes increasingly dissimilar to the in-domain data. For example, in \tbl{utt_baseline}, the AUC on TED increased from 33.57\% to 71.24\% whereas the AUC on SWBD only increased from 34.91\% to 52.31\%. The EER on TED is reduced by more than half whereas the EER on SWBD is only reduced by 23\% relatively. This is expected as the confidence module is trained only on in-domain data.

\presec
\subsection{Out-of-Domain Information for Confidence Estimation}
\postsec
Based on the previous observations, we explore various techniques that utilise the unlabelled OOD data to improve the confidence estimator while keeping the in-domain performance unchanged. As described in \sectdot{ood}, the pseudo transcriptions on the OOD data can be used to train confidence estimators and additional features from an LM trained on OOD text may also be useful. The OOD data with pseudo transcriptions was mixed with in-domain data in a 1:9 ratio for each minibatch. \tbl{improving_ted} shows the results when incorporating this additional OOD information when training a CEM and an R-EBM on the TED-LIUM dataset. Compared with the first row of \tbl{improving_ted} (\ie CEM and R-EBM baselines in \twotbl{word_baseline}{utt_baseline}), using pseudo transcriptions can effectively improve AUC and reduce EER. Although the improvement brought by additional OOD LM features is smaller, using both the pseudo transcription and OOD LM features yields the best confidence estimator. A similar observation is also made on the SWBD dataset. Therefore, both pieces of OOD information will be used for the following experiments.
\begin{table}[ht]
    \centering
    \begin{tabular}{ccccccc}
        \toprule
        \multirow{2}{*}{pseudo} & \multirow{2}{*}{LM} & \multicolumn{2}{c}{word-level} && \multicolumn{2}{c}{utterance-level} \\
        \cmidrule{3-4}\cmidrule{6-7}
         & & AUC$\uparrow$ & EER$\downarrow$ && AUC$\uparrow$ & EER$\downarrow$ \\
        \midrule
                     &              & 98.78 & 18.28 & & 71.24 & 23.17\\
        \midrule
        $\checkmark$ &              & \bf{98.85} & 16.41 & & 74.89 & 20.05\\
                     & $\checkmark$ & 98.73 & 18.56 & & 73.51 & 21.97\\
        $\checkmark$ & $\checkmark$ & \bf{98.85} & \bf{16.08} & & \bf{75.70} & \bf{19.21}\\
        \bottomrule
    \end{tabular}
    \caption{Word and utterance-level confidence estimation performance in AUC (\%) and EER (\%) on TED dataset with additional OOD information for CEM and R-EBM.}
    \label{tab:improving_ted}
    \posttbl
\end{table}

To further validate the effectiveness of the two proposed approaches, the OOD acoustic data is included during pretraining. This is a more challenging setup because confidence performance baseline will improve as pretraining can effectively reduce the WERs on OOD data. After continuing training the wav2vec 2.0 model with a mixture of in-domain and OOD data with a reduced learning rate, the new encoder is then fine-tuned on the LibriSpeech 100h dataset. The final WERs on the in-domain dataset are very similar to the baseline ASR model (within $\pm$0.1\%), but the WER is reduced by 10.6\% relative on TED-LIUM and 14.5\% relative on SWBD. An improved ASR model generally suggests that the quality of confidence scores are also improved. By comparing the confidence metrics before and after pretraining with OOD data shown in \tbl{improving_two}, AUCs are generally better with pretraining. However, EERs can be higher for the better ASR model because EER only represents a single operating point whereas AUC presents the overall picture of the confidence estimator at all operating points. Nevertheless, after including the OOD information for CEM or R-EBM, the confidence quality is consistently improved on OOD datasets, even when OOD acoustic data is used for pretraining.
\begin{table*}[ht!]
    \centering
    \begin{tabular}{lcrccccccrccccc}
        \toprule
        \multirow{3}{*}{dataset} & \multirow{3}{*}{OOD info} & \multicolumn{6}{c}{w/o OOD pretraining} && \multicolumn{6}{c}{w/ OOD pretraining}\\
        \cmidrule{3-8}\cmidrule{10-15}
        & & \multirow{2}{*}{WER} & \multicolumn{2}{c}{word-level} && \multicolumn{2}{c}{utterance-level} && \multirow{2}{*}{WER} & \multicolumn{2}{c}{word-level} && \multicolumn{2}{c}{utterance-level} \\
        \cmidrule{4-5}\cmidrule{7-8}\cmidrule{11-12}\cmidrule{14-15}
         & & & AUC$\uparrow$ & EER$\downarrow$ && AUC$\uparrow$ & EER$\downarrow$ & & & AUC$\uparrow$ & EER$\downarrow$ && AUC$\uparrow$ & EER$\downarrow$ \\
        \midrule
        \multirow{2}{*}{TED (test)} & & \multirow{2}{*}{9.4} & 98.78 & 18.28 & & 71.24 & 23.17 & & \multirow{2}{*}{8.4} & 98.87 & 17.55 & & 74.00 & 22.48\\
         & $\checkmark$ & & \bf{98.85} & \bf{16.08} & & \bf{75.70} & \bf{19.21} & & & \bf{99.06} & \bf{15.56} & & \bf{75.90} & \bf{20.37} \\
        \midrule
        \multirow{2}{*}{SWBD (RT03)} & & \multirow{2}{*}{28.3} & 96.45 & 18.59 & & 52.31 & 28.35 & & \multirow{2}{*}{24.2} & 96.80 & 19.00 & & 58.54 & 25.59\\
         & $\checkmark$ & & \bf{97.46} & \bf{14.79} & & \bf{60.95} & \bf{22.77} & & & \bf{97.54} & \bf{16.83} & & \bf{61.75} & \bf{23.07}\\
        \bottomrule
    \end{tabular}
    \caption{Confidence metrics in AUC (\%) and EER (\%) after using CEM \& R-EBM with additional OOD information on TED-LIUM \& SWBD. After pretraining with OOD data, ASR models have nearly the same performance on in-domain data but lower WERs on OOD data. This is a more challenging setup for making improvements on confidence estimators as the encoder has been exposed to OOD data.}
    \label{tab:improving_two}
    \posttbl
    \posttbl
\end{table*}

\presec
\subsection{Word-Level Confidence Calibration}
\postsec
Previous confidence metrics such as AUC and EER are only influenced by the rank ordering of the confidence scores, but not their absolute values. However, well calibrated word-level confidence scores can be important for some downstream applications. In other words, the absolute value of the confidence score should ideally reflect the probability of the word being recognised correctly. Two commonly used metrics for evaluating calibration performance is normalised cross-entropy (NCE)~\cite{Siu1997ImprovedEE} and expected calibration error (ECE)~\cite{Guo2017OnCO}. For a dataset with $N$ words, each word has a predicted confidence $p$ and a binary target $c$. NCE is defined as
\begin{equation}
    \text{NCE}(\mathbf{c},\mathbf{p}) = \dfrac{H(\mathbf{c}) - H(\mathbf{c},\mathbf{p})}{H(\mathbf{c)}},
\end{equation}
where $H(\mathbf{c})$ is the entropy of the target sequence and $H(\mathbf{c},\mathbf{p})$ is the binary cross-entropy between the target and the estimated confidence scores. When confidence estimation is systematically better than the word correct ratio ($\sum \mathbf{c}/N$), NCE is positive. For perfect confidence scores, NCE is 1. NCE measures how close the confidence score is to the the probability of the recognised word being correct. ECE is computed as the averaged gap between expected confidence and predicted confidence after binning all words into $M$ buckets, \ie 
\begin{equation}
    \text{ECE} = \sum_{m=1}^M\dfrac{|B_m|}{N}\bigg|\text{acc}(B_m) - \text{conf}(B_m)\bigg|,
\end{equation}
where $B_m$ represents the words fall within $m$-th bin ranked by their confidence scores. 
\begin{table}[ht]
    \centering
    \begin{tabular}{llcc}
        \toprule
        dataset & estimator & NCE$\uparrow$ & ECE$\downarrow$\\
        \midrule
        \multirow{3}{*}{TED (test)} & softmax & 0.1675 & 6.16 \\
         & CEM & 0.1221 & 3.39 \\
         & \;\;+OOD info & \bf{0.3473} & \bf{1.94} \\
        \midrule
        \multirow{3}{*}{SWBD (RT03)} & softmax & 0.2642 & 6.08 \\
         & CEM & 0.3377 & 1.33 \\
         & \;\;+OOD info & \bf{0.4694} & \bf{0.82} \\
        \bottomrule
    \end{tabular}
    \caption{Word-level NCE and ECE (\%) on TED \& SWBD test sets.}
    \label{tab:nce}
    \posttbl
\end{table}

Before computing NCE and ECE values, a piece-wise linear mapping (PWLM)~\cite{Evermann2000PosteriorPD,Guo2017OnCO} is estimated on the dev set and then applied to the test set. In this experiment, the PWLM uses 5 linear segments and 50 bins, and the computation of ECE also uses 50 bins. As shown in \tbl{nce}, on both OOD datasets, the word-level NCE and ECE are considerably better than the softmax or the CEM baseline after using OOD information during the training of CEM.

\presec
\subsection{Utterance-Level Data Selection}
\postsec
Although aggregating all word-level confidence scores can yield an utterance-level score, \cite{Li2021ResidualEM} showed that the utterance-level confidence is more effectively modelled by R-EBM which directly optimises the utterance-level objective. The improved utterance-level confidence scores can be readily used for data selection tasks. For active learning, utterances with low confidence are normally selected for manual transcription. For semi-supervised learning, utterances with high confidence are often included as additional training data because the hypotheses can be used as high-quality pseudo transcriptions.
\begin{table}[ht]
    \centering
    \begin{tabular}{llcc}
        \toprule
        dataset & estimator & bottom SER$\uparrow$ & top SER$\downarrow$\\
        \midrule
        \multirow{3}{*}{TED (test)} & softmax & \phantom{0}93.5 & 48.5 \\
         & R-EBM & \phantom{0}98.0 & 25.5 \\
         & \;\;+OOD info & \phantom{0}\bf{99.5} & \bf{18.5} \\
        \midrule
        \multirow{3}{*}{SWBD (RT03)} & softmax & \bf{100.0} & 57.5 \\
         & R-EBM & \phantom{0}99.5 & 23.0 \\
         & \;\;+OOD info & \bf{100.0} & \bf{17.0} \\
        \bottomrule
    \end{tabular}
    \caption{Sentence error rates (SERs) (\%) of the bottom and top 200 utterances of TED \& SWBD test sets.}
    \label{tab:ser}
    \posttbl
\end{table}

In \tbl{ser}, the SERs of 200 utterances with the lowest and the highest confidence scores from each dataset are reported. As expected, the R-EBM with additional OOD information during training can effectively filter utterances at either regime. Especially for the high-confidence utterances, OOD information is very helpful in reducing the SERs for R-EBM.

%% file: text/conclusion.tex
\presec
\section{Conclusion}
\postsec
\label{sec:conclusion}
Although model-based confidence scores are more reliable than softmax probabilities from end-to-end models as confidence estimators for both in-domain and OOD data, the performance on OOD data lags far behind the in-domain scenario. Using pseudo transcriptions to provide binary targets for training model-based confidence estimators and including additional features from an OOD LM are useful for improving the confidence scores on OOD datasets, even when pretraining uses OOD data. By exposing the CEM or R-EBM to OOD data, the calibration performance is also improved. Selecting OOD data using the improved confidence estimators is expected to considerably aid active or semi-supervised learning.

%% file: main.bbl
\begin{thebibliography}{10}

\bibitem{Wessel2001ConfidenceMF}
F.~Wessel, R.~Schl{\"u}ter, K.~Macherey, \& H.~Ney,
\newblock ``Confidence measures for large vocabulary continuous speech
  recognition,''
\newblock {\em IEEE Trans. on Speech and Audio Processing}, vol. 9, 2001.

\bibitem{Jiang2005ConfidenceMF}
H.~Jiang,
\newblock ``Confidence measures for speech recognition: A survey,''
\newblock {\em Speech Communication}, vol. 45, 2005.

\bibitem{Yu2011CalibrationOC}
D.~Yu, J.~Li, \& L.~Deng,
\newblock ``Calibration of confidence measures in speech recognition,''
\newblock {\em IEEE Trans. on Audio, Speech, and Language Processing}, vol. 19,
  2011.

\bibitem{Chan2004ImprovingBN}
R.H.Y.~Chan \& P.C.~Woodland,
\newblock ``Improving broadcast news transcription by lightly supervised
  discriminative training,''
\newblock {\em Proc. }{\em ICASSP}, Montreal, 2004.

\bibitem{Tr2005CombiningAA}
G.~T{\"u}r, D.Z.~Hakkani-T{\"u}r, \& R.~Schapire,
\newblock ``Combining active and semi-supervised learning for spoken language
  understanding,''
\newblock {\em Speech Communication}, vol. 45, 2005.

\bibitem{Riccardi2005ActiveLT}
G.~Riccardi \& D.~Hakkani-T{\"u}r,
\newblock ``Active learning: Theory and applications to automatic speech
  recognition,''
\newblock {\em IEEE Trans. on Speech and Audio Processing}, vol. 13, 2005.

\bibitem{Evermann2000PosteriorPD}
G.~Evermann \& P.C.~Woodland,
\newblock ``Posterior probability decoding, confidence estimation and system
  combination,''
\newblock {\em Proc. }{\em NIST Speech Transcription Workshop}, College Park,
  2000.

\bibitem{Mangu2000FindingCI}
L.~Mangu, E.~Brill, \& A.~Stolcke,
\newblock ``Finding consensus in speech recognition: Word error minimization
  and other applications of confusion networks,''
\newblock {\em Computer Speech \& Language}, 2000.

\bibitem{Seigel2011CombiningIS}
M.S.~Seigel \& P.C.~Woodland,
\newblock ``Combining information sources for confidence estimation with CRF
  models,''
\newblock {\em Proc. }{\em Interspeech}, Florence, 2011.

\bibitem{Kalgaonkar2015EstimatingCS}
K.~Kalgaonkar, C.~Liu, Y.~Gong, \& K.~Yao,
\newblock ``Estimating confidence scores on ASR results using recurrent neural
  networks,''
\newblock {\em Proc. }{\em ICASSP}, Brisbane, 2015.

\bibitem{DelAgua2018SpeakerAdaptedCM}
M.{\'A}.~Del-Agua, A.~Gim{\'e}nez, A.~Sanch{\'i}s, J.C.~Saiz, \& A.~Juan,
\newblock ``Speaker-adapted confidence measures for ASR using deep
  bidirectional recurrent neural networks,''
\newblock {\em IEEE/ACM Trans. on Audio, Speech, and Language Processing}, vol.
  26, 2018.

\bibitem{Ragni2018ConfidenceEA}
A.~Ragni, Q.~Li, M.J.F.~Gales, \& Y.~Wang,
\newblock ``Confidence estimation and deletion prediction using bidirectional
  recurrent neural networks,''
\newblock {\em Proc. }{\em SLT}, Athens, 2018.

\bibitem{Li2019BidirectionalLR}
Q.~Li, P.~Ness, A.~Ragni, \& M.J.F.~Gales,
\newblock ``Bi-directional lattice recurrent neural networks for confidence
  estimation,''
\newblock {\em Proc. }{\em ICASSP}, Brighton, 2019.

\bibitem{Li2021ABA}
B.~Li, A.~Gulati, J.~Yu, T.~Sainath, C.~Chiu, \etal,
\newblock ``A better and faster end-to-end model for streaming ASR,''
\newblock {\em Proc. }{\em ICASSP}, Toronto, 2021.

\bibitem{Tuske2021OnTL}
Z.~Tuske, G.~Saon, \& B.~Kingsbury,
\newblock ``On the limit of English conversational speech recognition,''
\newblock {\em Proc. }{\em Interspeech}, Brno, 2021.

\bibitem{Li2020DevelopingRM}
J.~Li, R.~Zhao, Z.~Meng, Y.~Liu, W.~Wei, \etal,
\newblock ``Developing RNN-T models surpassing high-performance hybrid models
  with customization capability,''
\newblock {\em Proc. }{\em Interspeech}, Shanghai, 2020.

\bibitem{Graves2012SequenceTW}
A.~Graves,
\newblock ``Sequence transduction with recurrent neural networks,''
\newblock {\em Proc. }{\em ICML Workshop on Representation Learning},
  Edinburgh, 2012.

\bibitem{Chorowski2015AttentionBasedMF}
J.~Chorowski, D.~Bahdanau, D.~Serdyuk, K.~Cho, \& Y.~Bengio,
\newblock ``Attention-based models for speech recognition,''
\newblock {\em Proc. }{\em NIPS}, Montreal, 2015.

\bibitem{Oneata2021AnEO}
D.~Oneata, A.~Caranica, A.~Stan, \& H.~Cucu,
\newblock ``An evaluation of word-level confidence estimation for end-to-end
  automatic speech recognition,''
\newblock {\em Proc. }{\em SLT}, Shenzhen, 2021.

\bibitem{Woodward2020ConfidenceMI}
A.~Woodward, C.~Bonn{\'i}n, I.~Masuda, D.~Varas, E.~Bou, \& J.C.~Riveiro,
\newblock ``Confidence measures in encoder-decoder models for speech
  recognition,''
\newblock {\em Proc. }{\em Interspeech}, Shanghai, 2020.

\bibitem{Li2021ConfidenceEF}
Q.~Li, D.~Qiu, Y.~Zhang, B.~Li, Y.~He, \etal,
\newblock ``Confidence estimation for attention-based sequence-to-sequence
  models for speech recognition,''
\newblock {\em Proc. }{\em ICASSP}, Toronto, 2021.

\bibitem{Qiu2021LearningWC}
D.~Qiu, Q.~Li, Y.~He, Y.~Zhang, B.~Li, \etal,
\newblock ``Learning word-level confidence for subword end-to-end ASR,''
\newblock {\em ICASSP}, 2021.

\bibitem{Kumar2020UtteranceCM}
A.~Kumar, S.~Singh, D.~Gowda, A.~Garg, S.~Singh, \& C.~Kim,
\newblock ``Utterance confidence measure for end-to-end speech recognition with
  applications to distributed speech recognition scenarios,''
\newblock {\em Proc. }{\em Interspeech}, Shanghai, 2020.

\bibitem{Li2021ResidualEM}
Q.~Li, Y.~Zhang, B.~Li, L.~Cao, \& P.C.~Woodland,
\newblock ``Residual energy-based models for end-to-end speech recognition,''
\newblock {\em Proc. }{\em Interspeech}, Brno, 2021.

\bibitem{Liu2021UtterancelevelNC}
W.~Liu \& T.~Lee,
\newblock ``Utterance-level neural confidence measure for end-to-end children
  speech recognition,''
\newblock {\em Proc. }{\em ASRU}, Cartagena, 2021.

\bibitem{Qiu2021MultiTaskLF}
D.~Qiu, Y.~He, Q.~Li, Y.~Zhang, L.~Cao, \& I.~McGraw,
\newblock ``Multi-task learning for end-to-end ASR word and utterance
  confidence with deletion prediction,''
\newblock {\em Proc. }{\em Interspeech}, Brno, 2021.

\bibitem{Ovadia2019CanYT}
Y.~Ovadia, E.~Fertig, J.~Ren, Z.~Nado, D.~Sculley, \etal,
\newblock ``Can you trust your model's uncertainty? Evaluating predictive
  uncertainty under dataset shift,''
\newblock {\em Proc. }{\em NeurIPS}, Vancouver, 2019.

\bibitem{Kahn2020LibriLightAB}
J.~Kahn, M.~Rivi{\`e}re, W.~Zheng, E.~Kharitonov, Q.~Xu, \etal,
\newblock ``Libri-light: A benchmark for ASR with limited or no supervision,''
\newblock {\em Proc. }{\em ICASSP}, Barcelona, 2020.

\bibitem{Panayotov2015LibrispeechAA}
V.~Panayotov, G.~Chen, D.~Povey, \& S.~Khudanpur,
\newblock ``LibriSpeech: An ASR corpus based on public domain audio books,''
\newblock {\em Proc. }{\em ICASSP}, Brisbane, 2015.

\bibitem{hernandez2018tedlium3}
F.~Hernandez, V.~Nguyen, S.~Ghannay, N.~Tomashenko, \& Y.~Est{\`e}ve,
\newblock ``TED-LIUM 3: Twice as much data and corpus repartition for
  experiments on speaker adaptation,''
\newblock {\em Proc. }{\em SPECOM}, Leipzig, 2018.

\bibitem{Rousseau2014EnhancingTT}
A.~Rousseau, P.~Del{\'e}glise, \& Y.~Est{\`e}ve,
\newblock ``Enhancing the TED-LIUM corpus with selected data for language
  modeling and more TED talks,''
\newblock {\em Proc. }{\em LREC}, Reykjavik, 2014.

\bibitem{Zhang2020PushingTL}
Y.~Zhang, J.~Qin, D.S.~Park, W.~Han, C.~Chiu, \etal,
\newblock ``Pushing the limits of semi-supervised learning for automatic speech
  recognition,''
\newblock {\em Proc. }{\em NeurIPS SAS Workshop}, Vancouver, 2020.

\bibitem{Baevski2020wav2vec2A}
A.~Baevski, H.~Zhou, A.r.~Mohamed, \& M.~Auli,
\newblock ``wav2vec 2.0: A framework for self-supervised learning of speech
  representations,''
\newblock {\em Proc. }{\em NeurIPS}, Vancouver, 2020.

\bibitem{Park2019SpecAugmentAS}
D.~Park, W.~Chan, Y.~Zhang, C.C.~Chiu, B.~Zoph, \etal,
\newblock ``SpecAugment: A simple data augmentation method for automatic speech
  recognition,''
\newblock {\em Proc. }{\em Interspeech}, Graz, 2019.

\bibitem{Schuster2012JapaneseAK}
M.~Schuster \& K.~Nakajima,
\newblock ``Japanese and Korean voice search,''
\newblock {\em Proc. }{\em ICASSP}, Kyoto, 2012.

\bibitem{Saito2015ThePP}
T.~Saito \& M.~Rehmsmeier,
\newblock ``The precision-recall plot is more informative than the ROC plot
  when evaluating binary classifiers on imbalanced datasets,''
\newblock {\em PLOS ONE}, vol. 10, 2015.

\bibitem{Siu1997ImprovedEE}
M.~Siu, H.~Gish, \& F.~Richardson,
\newblock ``Improved estimation, evaluation and applications of confidence
  measures for speech recognition,''
\newblock {\em Proc. }{\em Eurospeech}, Rhodes, 1997.

\bibitem{Guo2017OnCO}
C.~Guo, G.~Pleiss, Y.~Sun, \& K.~Weinberger,
\newblock ``On calibration of modern neural networks,''
\newblock {\em Proc. }{\em ICML}, Sydney, 2017.

\end{thebibliography}
